\documentclass[amssymb,aps]{revtex4}

\begin{document}
\title{A modified orbital motion limited (OML) theory}
\author{Zahida Ehsan$^{(a,b)}\thanks{%
For correspondence: ehsan.zahida@gmail.com},$ N. L.
Tsintsadze$^{(c)}$, and S. Poedts$^{(a)}$}
\affiliation{
(a)~Center for Plasma Astrophysics, K.U.Leuven, Celestijnenlaan 200B, 3001 Leuven, Belgium\\
(b)~National Center for Physics, Quaid-i-Azam University Campus, Islamabad, Pakistan\\
(c)~E. Andronikashvili Institute of Physics,0171 Georgia}

\begin{abstract}
The validity of the orbital motion limited (OML) theory is reviewed with
reference to the floating potential acquired by a spherical object immersed
in a plasma. A new and perhaps more realistic approach for obtaining the
floating potential is introduced by including the current outward from the
spherical object and the current coming from infinity. This novel approach
is also valid for cases where the standard OML theory ceases to apply.
\end{abstract}

\pacs{52.27.Lw; 52.35.Fp; 52.35.Mw}

\maketitle

 The collection of plasma particles by an ideal spherical
object has applications e.g., in the operation of Langmuir probes,
spacecraft and other immersed objects and, especially, for grains in
dusty plasma\cite {1,2,3,4,5,6,7,8}. Since the different objects in
the mentioned phenomena behave in a similar way, we intend to use
the general term `object' throughout this paper.

There have been many theoretical attempts to solve the problem of the
charging of objects immersed in plasmas starting from Mott-Smith and Langmuir%
\cite{9}, and followed by many other authors\cite
{10,11,12,13,14,15,16,17,18,19,20}. One needs to determine the floating
potential to study the objects introduced into the plasma, that is the
potential at which ion and electron currents to the object are balanced.
However, there is no universally accepted theory for the quantitative
description of the particle charging in plasmas. Yet, owing to its
simplicity, the orbital motion limited (OML) approach is often applied for
spherical objects in low density plasmas\cite{9,12}. This approach deals
with collisionless electron and ion trajectories in the vicinity of a small
object and allows us to determine the cross-sections for electron and ion
collection from the conservation laws of energy and angular momentum. This
seems to be quite a reasonable approach, for example in the case of dusty
plasmas, where the mean free paths of the ion collisions with atoms are
usually very long compared to the Debye length such that $a<<\lambda
_{d}<<l_{i(e)}$\cite{17}, where $a$\ and $\lambda _{d}$\ are the radius of
object and plasma screening length (the corresponding Debye radius),
respectively, and $l_{i(e)}$\ denotes the mean free path of the ions
(electrons). It is also assumed that the dust particle is rather isolated in
the sense that other dust particles do not affect the motion of electrons
and ions in its vicinity. However, the validity of the OML approach is
questionable. Strictly speaking, the OML approach is valid only when there
is no absorption radius, which is the absorption radius imagined as a
theoretical spherical surface where the grazing orbit is the limiting orbit
that an ion of a particular energy will hit the surface of object. For any
distribution function containing slow ions, the absence of the absorption
radius can be reformulated in a condition on the shielding potential around
the body, which must decrease more slowly than $1/r^{2}$. Allen, Annaratone
and de Angelis have shown that the OML approach is not congruent relevant to
dusty plasma phenomena\cite{21}. However, later on it was proved that with
limits of a vanishing body radius, OML theory is consistent for cases
relevant to dusty plasmas, and thus can be used as a useful approximation
for quantitative calculations.\cite{18}.

The purpose of this letter is to raise pertinent questions on the existing
OML theory and to present a new approach for obtaining the floating
potential which is fundamental in understanding to occurring phenomenon.

Consider the well-known OML result that $\varphi _{s}$ is a solution of
\begin{equation}
\left( \frac{T_{e}}{m_{e}}\right) ^{1/2}\exp \left( -\frac{e|\varphi _{s}|}{%
K_{B}T_{e}}\right) =\left( \frac{T_{i}}{m_{i}}\right) ^{1/2}\left( 1+\frac{%
e|\varphi _{s}|}{K_{B}T_{i}}\right),  \eqnum{1}
\end{equation}
where $m$ is the particle mass, $T$ is the temperature, and $K_{B}$ is
Boltzmanns constant. Here, we adopt the conventions that subscripts $e$ and $%
i$ denote properties of the plasma electrons and ions respectively. $e$ is
the magnitude of the electron charge and $\varphi _{s}$ is the surface
potential of the object. The above equation (1) seems not to be general as
it is always useful to consider a specific situation, for example when $%
e|\varphi _{s}|>K_{B}T_{e}$, and on the other hand when the object is
removed from a plasma i.e., $\varphi _{s}=0,$ it becomes invalid. Yet, it
yields a good approximation and provides a good description on the surface
of the object as it assumes neutrality condition ($n_{io}\backsimeq n_{eo})$
.

Here, we suggest another derivation of the above equation for the total
current since the floating condition is
\begin{equation}
I_{i}+I_{e}=0, \eqnum{2}
\end{equation}
where $I_{i}$ and $I_{e}$ are the ion and electron currents collected by the
body. For the case when there is no object in the plasma, $I_{i}=0$ and $%
I_{e}=0.$ For a spherical object with some potential around it, there can be
two types of current influenced by this potential, one is $I_{\alpha }^{-}$ (%
$\alpha $ for the electron and ion species) which is outward from the object
and $I_{\alpha }^{+}(\varphi _{s}=0)$ which comes from infinity. It is
important to note that $I_{\alpha }^{-}$\ is is not due to the reflection of
incoming electrons (or ions) from the object's surface\cite{22,23,24}. Fig.
(1) illustrates the ion and electron currents to and from the spherical
object of radius $a$. Effects of other currents like thermionic, secondary
or photoemission, etc.\cite{2,25}. are ignored here and the total current is
given by
\begin{equation}
I_{\alpha }^{total}=I_{\alpha }^{-}(\varphi _{s})+I_{\alpha
}^{+}(\varphi _{s}=0).  \eqnum{3}
\end{equation}
If no object is present $(\varphi _{s}=0)$, $I_{\alpha
}^{total}=I_{\alpha }^{-}+I_{\alpha }^{+}=0$, providing $I_{\alpha
}^{+}=-I_{\alpha }^{-}$. Now, using the OML approximation for the
electrons, we can write
\begin{equation}
I_{e}=-4\pi r_{o}^{2}n_{oe}e\left( \frac{m_{e}}{2\pi
K_{B}T_{e}}\right)^{3/2}\left\{
\int_{0}^{\infty}v_{z}e^{-\frac{m_{e}v^{2}}{2\pi K_{B}T_{e}}}\;d{v} + 
\int_{-\infty}^{0}v_{z}e^{-\frac{m_{e}v^{2}+2e|\varphi
_{s}|}{2K_{B}T_{e}}}\;d{v}
\right\}.   \eqnum{4}
\end{equation}
Integration of the above equation gives us the electron current
\begin{equation}
I_{e}=-4\pi r_{o}^{2}n_{oe}e\left( \frac{K_{B}T_{e}}{2\pi
m_{e}}\right) ^{1/2}\left( 1-e^{-\frac{e|\varphi _{s}|}{K_{B}T_{e}}}
\right).  \eqnum{5}
\end{equation}
Now we consider an ion approaching an object from the bulk plasma at the
edge of its shielding cloud $(r=\infty )$ at speed $v_{i}$ and {\bf striking}
the spherical object with surface potential $\varphi _{s}$ and speed $%
v_{i}^{^{\prime }}$. We assume that there are no collisions with other
particles in the shielding cloud. The conservation of energy and angular
momentum is expressed by the following equations
\begin{equation}
\frac{m_{i}v_{i}^{2}}{2}=\frac{mv_{i}^{^{\prime }2}}{2}-e|\varphi _{s}|,
\eqnum{6}
\end{equation}
where from above $v_{i}^{^{\prime }}=v_{i}\left( 1+\frac{2e|\varphi _{s}|}{%
m_{i}v_{i}^{2}}\right) ^{1/2}$, and
\begin{equation}
m_{i}v_{i}h_{c}=m_{i}v_{i}^{^{\prime }}a,  \eqnum{7}
\end{equation}
where we have defined a critical impact parameter $h_{c}$ for which ions
have a grazing collision with the object which is
\begin{equation}
h_{c}=a\left( 1+\frac{2e|\varphi _{s}|}{m_{i}v_{i}^{2}}\right)
^{1/2}. \eqnum{8}
\end{equation}
The cross section for the collection of ions is, therefore,
\begin{equation}
\sigma _{i}=\pi h_{c}^{2}=\pi a^{2}\left( 1+\frac{2e|\varphi _{s}|}{%
m_{i}v_{i}^{2}}\right) =4\pi r_{o}^{2}\left( 1+\frac{2e|\varphi _{s}|}{%
m_{i}v_{i}^{2}}\right).  \eqnum{9}
\end{equation}
For a given Maxwellian distribution in the bulk plasma, the ion flux to the
object surface is determined by the integration of the corresponding
cross-section with $f_{i}(v_{i})$:
\begin{equation}
I_{i}=n_{io}\int_{0}^{\infty }v\sigma _{i}f_{i}(v)d^{3}v.  \eqnum{10}
\end{equation}
Hence, the ion current $\,I_{i}^{+}$ to the grain reads
\begin{equation}
I_{i}^{+}=Z_{i}e\left( \frac{m_{i}}{2\pi K_{B}T_{i}}\right)
^{3/2}\int_{0}^{\infty }v\sigma _{i}e^{
-\frac{m_{i}v^{2}}{2K_{B}T_{i}}}\;d{v}.  \eqnum{11}
\end{equation}
For spherical coordinates, the integration in Eq.~(11) performed with the
use of the Eq.~(10) gives
\begin{equation}
I_{i}^{+}=4\pi r_{o}^{2}Z_{i}n_{oi}e\left( \frac{K_{B}T_{i}}{2\pi m_{i}}%
\right) ^{1/2}\left( 1+\frac{e|\varphi _{s}|}{K_{B}T_{i}}\right).
\eqnum{12}
\end{equation}
Similarly, for the outward current $I_{i}^{-}$ we have
\begin{equation}
I_{i}^{-}=4\pi r_{o}^{2}Z_{i}n_{oi}e\left( \frac{m_{i}}{2\pi K_{B}T_{i}}%
\right) ^{3/2}\int_{-\infty }^{0}v_{z}
e^{-\frac{m_{i}v^{2}}{2K_{B}T_{i}}}\;d{v}. \eqnum{13}
\end{equation}
Hence, we get
\begin{equation}
I_{i}^{-}=-4\pi r_{o}^{2}Z_{i}en_{oi}\left( \frac{K_{B}T_{i}}{2\pi m_{i}}%
\right) ^{1/2}.  \eqnum{14}
\end{equation}
Finally, the total ion current reads
\begin{equation}
I_{i}=I_{i}^{+}+I_{i}^{-}=4\pi r_{o}^{2}Z_{i}en_{oi}\left( \frac{K_{B}T_{i}}{%
2\pi m_{i}}\right) ^{1/2}\frac{e|\varphi _{s}|}{K_{B}T_{i}}.  \eqnum{15}
\end{equation}
Now we can determine the floating potential of the object using the results
from above Eq.~(15) and equating to the electron current Eq.~(5).
\begin{equation}
\left( \frac{T_{e}}{m_{e}}\right) ^{1/2}\left[ 1-\exp \left( -\frac{%
e|\varphi _{s}|}{K_{B}T_{e}}\right) \right] =\left( \frac{T_{i}}{m_{i}}%
\right) ^{1/2}\left( \frac{e|\varphi _{s}|}{K_{B}T_{i}}\right).
\eqnum{16}
\end{equation}
This new equation (16) is also valid for $|\varphi _{s}|=0$. It also
represents that when $\beta =T_{i}/T_{e}$\ increases the normalized
potential $\Phi =$\ $\frac{e|\varphi _{s}|}{K_{B}T_{e}}$\ will also increase
as can be seen in Fig. (2). However, one question arises about the equality $%
n_{oe}$ $\backsim $ $Z_{i}n_{oi}$, since near the surface, due to the
collection of ions, the plasma may be inhomogeneous, or $Z_{i}n_{oi}>n_{oe}$%
. In addition to this, ions can also lose energy in rare collisions with
atoms and become trapped in finite orbits by the electric field of a charged
particle\cite{26,27,28}. Trapped ions effect was studied using molecular
dynamics calculations \cite{28}and with the help of analytical methods.\cite
{29,30}. In these papers, it was shown that the density of the trapped ions
(with negative total energy) could be greater than the density of the free
ions (with positive total energy) in the vicinity of a charged dust particle
and thus played an important role in the screening of the particle. How many
ions can also be trapped near the object in collisionless plasmas and what
is their density distribution is still an unsolved question\cite{21},
however in future we intend to include the trapped ions effect in modified
theory for the accuracy of model.

\subsection*{Acknowledgments}

The results presented here are obtained in the framework of GOA
project 2009-009 (K.U.Leuven), the European Commissions Seventh
Framework Program (FP7/2007- 2013) under the grant agreement no.
218816 (SOTERIA project, www.soteria-space.eu), Georgian Science
Foundation Grant Project No. 1-4/16(GNSF/ST09 305 4-140 and Higher
Education Commission of Pakistan.

\subsection*{Figure captions: }

Fig.~(1): Ions and electrons current to and from the spherical
object.

Fig.~(2): Normalized potential $\Phi =$\ $e|\varphi _{s}|K_{B}T_{e}$ versus $%
\beta =T_{i}/T_{e}$\ .

\end{document}